\documentclass{icrc}

\usepackage{times}
\usepackage{graphicx} 
\begin{document}

\title{ Precise parametrizations of muon energy losses in water}
\author[1]{S. Klimushin}
\author[1]{E. Bugaev}
\affil[1]{
Institute for Nuclear Research, Russian Academy of Science, Moscow 117312, Russia}
\author[1,2]{I. Sokalski}
\affil[2]{now at DAPNIA/SPP, CEA/Saclay, 91191 Gif-sur-Yvette CEDEX, France}
\correspondence{klim@pcbai11.inr.ruhep.ru}

\firstpage{1}
\pubyear{2001}


\maketitle

\begin{abstract}
The description of muon propagation through large depths of matter,
based on a concept of the correction factor, is proposed.
The results of Monte-Carlo calculations of this correction
factor are presented. The parametrizations for continuous energy
loss coefficients, valid in the broad interval of muon energies,
and for the correction factor are given. The concrete calculations 
for pure water are presented.
\end{abstract}

\section{Introduction}
The existing methods of calculations of muon propagation through thick layers of matter 
(basically for a standard rock) are largely presented in the literature.
There are two completely different approaches in such calculations. In first (historically) works on
the subject of muon propagation the integro-differential kinetic equation for muon flux had been formulated and
approximately solved using semianalytical methods. We mention here one of the pioneering works of this approach:
~\citep{ZM62}, and one of the most recent papers:~\citep{NSB94} (this latter paper contains also the large bibliography).
Works of the second approach use Monte Carlo (MC) technique for propagation studies: the main element in
such works is simulation of sequences of free flights and interactions during muon passage through a medium. 
One of the most early works using this method is~\citep{HPW63}, the most recent one is the work of the present authors
~\citep{MUM1}, containing also many bibliographic references.

\section{Method}
The integral flux of muons in matter with energies above cut-off $E_f$
expected at slant depth $R$ at a zenith angle of $\theta$ taking into account 
{\it fluctuating} character of muon losses is conventionally described by
\begin{equation}\label{Int_fl}
I_{fl}(\geq E_f,R,\theta)=\int_0^{\infty}{P(E_0,R,\geq E_f)D(E_0,\theta)\,dE_0},
\end{equation}
where $P(E_0,R,\geq E_f)$ is the probability that a muon, having starting energy $E_0$,
after passing of path $R$ will survive with final energy above cut-off $E_f$
and $D(E_0,\theta)=dN(E_0,\theta)/dE_0$ is a sea level differential angular spectrum.
In a case of a flat surface the slant depth is expressed by $R=h/\cos\theta$, where $h$
is a vertical depth below the surface.

With the assumption of {\it continuous} energy loss rate of muon in matter, $L(E)=-dE/dx=a(E)+b(E)E$, the 
integral flux $I_{cl}$ is derived from the equation 
\begin{equation}\label{Int_cont}
I_{cl}(\geq E_f,R,\theta)=\int_{E_s}^{\infty}{D(E_0,\theta)\,dE_0},
\end{equation}
where the low limit of the integral, $E_s$, is the value of starting energy $E_0$ which results, 
after passing of path $R$, in the final energy $E_f$. This value is derived from the solution of the integral equation
\begin{equation}\label{Ra}
-\int_{E_s}^{E_f} \frac{dE}{L(E)}=R.
\end{equation}

The method proposed by us for a calculation of real integral muon flux allowing for loss fluctuations consists in the following.
As is shown in ~\citep{KBS}, if we approximate the muon energy loss function $L(E)$ by a linear energy dependence,
\begin{equation}\label{Leff}
                               L(E)=\alpha + \beta E
\end{equation}
(where $\alpha$ and $\beta$ are independent on energy), and if we parametrize the
sea level muon spectrum by dependencies of the type 
$$
D(E_0) \propto \frac {E_{0}^{-\gamma}} {1+E_{0}/\varepsilon_{0}},
$$
then the integral flux $I_{cl}$ (Eq.\ref{Ion}) can be expressed analytically. Further, to calculate a real
$I_{fl}$ we must know the {\it correction factor} defined by the ratio 
\begin{equation}\label{Cf}
C_{f}(\geq E_f,R,\theta)=\frac {I_{cl}(\geq E_f,R,\theta)} {I_{fl}(\geq E_f,R,\theta)}.  
\end{equation}
In principle, this factor can be calculated using known codes for muon propagation through matter. 
In this work we apply for this aim the MUM code
described in previous work of the present authors~\citep{MUM1}. 
As we will see, this factor, as a function of muon energy and depth, can be easily parametrized 
by simple expressions. As a result, the real flux $I_{fl}$ is conveniently
expressed by the formula, containing continuous energy loss coefficients $\alpha$ and $\beta$. 
\begin{figure}[t]
\vspace*{2.0mm} 
\includegraphics[width=8.3cm]{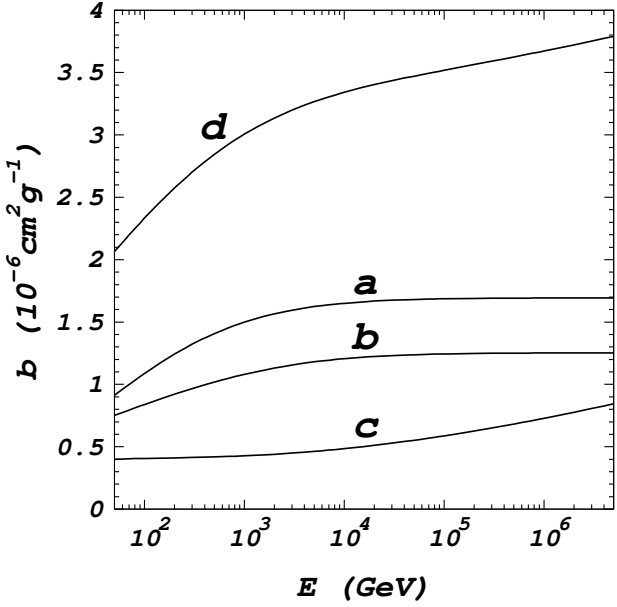} 
\protect\caption{
   The b-terms of muon radiative energy losses in pure water (as MUM code output). Losses due to  
   $e^{+}e^{-}$~pair production $b_p$ (curve (a)), bremsstrahlung $b_b$ (b), and photonuclear 
   interaction $b_n$ (c) are shown. Curve (d) is the sum of all b-terms.  
\label{fig:Btermf}}
\end{figure}
\begin{figure}[t]
\vspace*{2.0mm} 
\includegraphics[width=8.3cm]{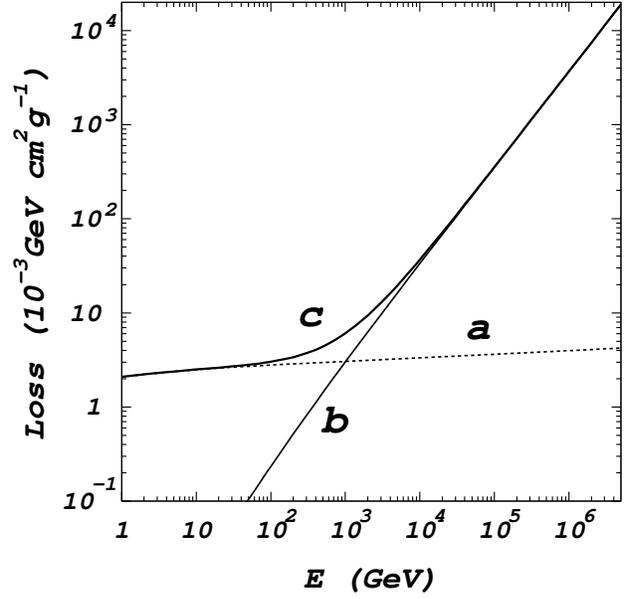} 
\protect\caption{
   Muon energy losses in pure water as a function of energy $E$ (as MUM code output). Curve (a) is loss due to ionization $a(E)$ and 
   (b) is total loss due to all radiative processes $b(E)E$. Curve (c) describes total energy losses $a(E)+b(E)E$.
\label{fig:Totlossf}}
\end{figure}
\section{Continuous energy losses}
\begin{table*}[tbh]
\protect\caption{ Coefficients $b_{ij}~(10^{-6}\mbox{cm}^2\mbox{g}^{-1})$ of the fitting formula~
                 (\protect\ref{Bterm}) for b-terms of muon energy losses in water. Maximum absolute values of relative errors are shown.  
\label{tab:bttab}}
\center{\begin{tabular}{lcrrrrrc} \hline \hline
b-term & subscript i &$b_{i0}~~~~~$ & $b_{i1}~~~~~$ & $b_{i2}~~~~~$ & $b_{i3}~~~~~$ & $b_{i4}$~~~~~& Max.err,$\%$ \\\hline
$e^{+}e^{-}$~pairs,~$b_p$ &p&$-11.31\times10^{-1} $ & $7.876 \times10^{-1}$ & $-8.192\times10^{-2}$ & $3.763\times10^{-3}$ & $-6.437\times10^{-5}$ & 0.8\\ 
bremsstrahlung,~$b_b$ &b    &$-1.149\times10^{-1}$ & $2.963 \times10^{-1}$ & $-2.165\times10^{-2}$ & $5.630\times10^{-4}$ & $-2.119\times10^{-6}$ & 0.9\\ 
photonuclear,  ~$b_n$ &n    &$ 3.903\times10^{-1}$ & $ 9.355\times10^{-3}$ & $-3.378\times10^{-3}$ & $4.913\times10^{-4}$ & $-1.216\times10^{-5}$ & 0.6 \\\hline \hline
\end{tabular}}
\end{table*} 
For the description of continuous energy losses of muon in water needed for solving the integral equation~(\ref{Ra})
we have done parametrizations based on output data from MUM code.
All formulas for cross sections used in MUM code for muon energy loss computations are described in details in Ref.~\citep{MUM1}.

The energy loss rate per unit of path $x$ is given, conventionally, by the expression
$L(E)=-dE/dx=a(E)+b(E)E$, where $a(E)$ is loss due to ionization and
$b(E)=b_p(E)+b_b(E)+b_n(E)$ is the sum of coefficients for all radiative processes: $e^{+}e^{-}$~pair production ($b_p$),
bremsstrahlung ($b_b$) and photonuclear interaction ($b_n$).   

Ionization loss was taken as composition of 2 processes
$a(E)=a_c(E)+a_e(E)$, where $a_c$ is classic ionization calculated using Bethe-Bloch formula
and $a_e$ results from $e$-diagrams for 
bremsstrahlung being treated as a part of ionization process following~\citep{bremkok}
($\gamma$-quantum is emitted by atomic electron). 
Taking into account the last process leads to $1.8\,\%,3.4\,\%,5.5\,\%$ increase of ionization loss
for 100 GeV, 1 TeV, 10 TeV muons, correspondingly.
The approximation formula for $a_c$ is given by 
\begin{eqnarray}\label{Ion}
a_c(E)=a_{c_0}+a_{c_1}\ln \left(\frac{W_{max}}{m_{\mu}}\right), \nonumber \\  
W_{max}=\frac{E}{1+{m^2_{\mu}}/{(2m_{e}E)}},         
\end{eqnarray}
where $W_{max}$ is maximum energy transferable to the electron and $m_\mu$, $m_e$
are the rest masses of muon and electron. 
The set of cofficients, in units of ($10^{-3}$GeVcm${}^2$g${}^{-1}$), 
\begin{eqnarray*}
            a_{c_0}=2.106,~a_{c_1}=0.0950~ \mbox{  for } E\leq 45 \mbox{ GeV,}\\
            a_{c_0}=2.163,~a_{c_1}=0.0853~ \mbox{  for } E>45\mbox{ GeV,} 
\end{eqnarray*}
gives the error for parametrization~(\ref{Ion}) smaller than 0.2$\,\%$ for (1--$10^8)\,$GeV range. 
For $a_e$ the following polynomial approximation ($E$ in units of (GeV)): 
\begin{eqnarray}\label{Ionbr}
a_e(E)=3.54 + 3.785\ln E + 1.15\ln^{2}E \nonumber \\ 
       + 0.0615\ln^{3}E \qquad (10^{-6}\mbox{GeVcm}^2\mbox{g}^{-1}),  
\end{eqnarray}
has the error $\leq 0.2\,\%$ for (50--$10^8)\,$GeV range.   
Finally, the sum of~(\ref{Ion}) and~(\ref{Ionbr}) has the error of $a(E)$ approximation 
smaller than 0.2$\,\%$ for (1--$10^8)\,$GeV range. 

Radiative energy losses of muon in water has been calculated using the cross sections from the following works: 
~\citep{brembb1}~for bremsstrahlung, \citep{pairkok} for $e^{+}e^{-}$~
pair production, and~\citep{phnubb} 
for photonuclear interaction. Fig.~\ref{fig:Btermf} sketches corresponding b-terms of radiative energy losses. 

In principle MUM code gives values for all b-terms up to $10^9$ GeV.  
Note that the logarithmic rise with an energy of the photoabsorption cross section, used in the model of Ref.~\citep{phnubb}
and resulting in corresponding increase of the $b_n$-term, is experimentally proved already up to the photon energy
$\sim 10^5$ GeV.

For parametrization of b-terms we have used the same functional form as in Ref.~\citep{brembb2} but increased
the power of polynomial to improve accuracy and enlarge the range of application up to 100 PeV without division
on energy subintervals ($E$ in units of (GeV)):
\begin{equation}\label{Bterm}
b_{i}(E)=\sum_{j=0}^4 b_{ij}\ln^{j}E, \qquad \mbox{where}~i=p,~b,~n.  
\end{equation}
Corresponding coefficients of this decomposition are collected in Tab.~\ref{tab:bttab}.
This fit works with typical errors $\pm$(0.2--0.4)$\%$ within (50--7$\times 10^7$) GeV for all b-terms.\\
The sum $b(E)$ of fits for b-terms is valid for the energy range (50--$10^8$) GeV with the relative error~$<0.5\,\%$.

Fig.~\ref{fig:Totlossf} shows muon energy loss rate in water. Note that the radiative losses account for approximately $5\,\%$ 
of the value of total losses at muon energy 70 GeV, $50\,\%$ at 1 TeV and dominate as $95\,\%$ at 20 TeV.
The sum of all parametrizations~(\ref{Ion}),~(\ref{Ionbr}) and~(\ref{Bterm}) results in the description 
of total energy losses $L(E)$ with excellent accuracy $<0.3\,\%$ with varying sign of error for 
energy range (1--$10^8$) GeV mainly because of b-term error compensations (even for region (1--50) GeV). 
Thanks to this fact just this parametrization of $L(E)$ has been used for the numerical computation of 
integral flux $I_{cl}$ defined by (\ref{Int_cont}) and (\ref{Ra}).

Further, we realized that the total losses $L(E)$ may be described with an accuracy better than $2.5\,\%$ 
(with varying sign of error) for energy range (3--3$\times 10^6$) GeV by 3-slope linear fit  
\begin{equation}\label{Totloss}
            L(E)=\alpha + \beta E, 
\end{equation}
where
\begin{eqnarray*} 
\alpha=\alpha_0=2.30, \beta=\beta_0=15.50 \mbox{ for~}  E\leq E_{01}=30.0 \mbox{GeV,}~~~~~\\
\alpha=\alpha_1=2.67, \beta=\beta_1=3.40 \qquad \qquad \qquad \qquad \qquad \qquad \\ 
\mbox{ for~} E_{01} <E \leq E_{12}=35.3 \mbox{TeV,} \qquad \\
\alpha=\alpha_2=-6.50,\beta=\beta_2=3.66  \mbox{ for~ } E > E_{12}=35.3 \mbox{TeV,}~~~~
\end{eqnarray*}
and $\alpha$ are in $(10^{-3}$GeVcm${}^2$g${}^{-1})$ and $\beta$ in $(10^{-6}\mbox{cm}^2\mbox{g}^{-1})$.
The energy losses expressed by this parametrizations have sense of effective ones, for example in energy region 30 GeV--35 TeV 
the values $\alpha=2.67$ and $\beta=3.40$ represent effective energy losses due to ionization and radiative processes, correspondingly.
The availability of the linear dependence~(\ref{Totloss}) leads 
to the possibility to derive an analytical formula for underwater integral flux (see~\citep{KBS1}).   

\section{Correction factor}
\begin{figure}[t]
\vspace*{2.0mm} 
\includegraphics[width=8.3cm]{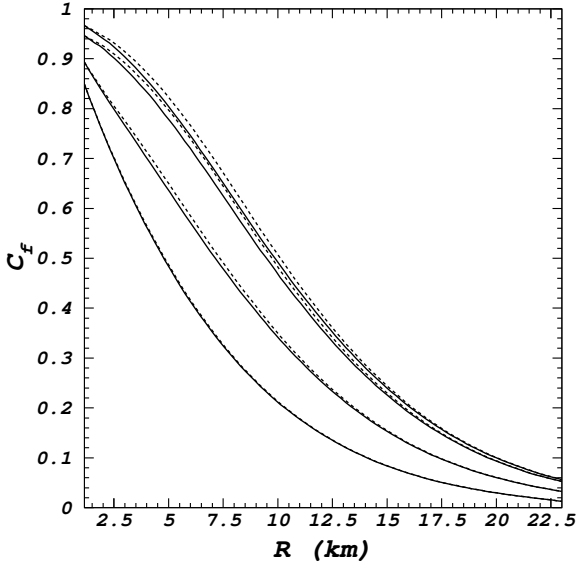} 
\protect\caption{ 
Correction factor $C_f$ as a function of slant depth $R$ in pure water. 
The results obtained using sea level spectrum defined by expression~(5) of Ref.~\citep{KBS1} are given.
Solid curves correspond to numerical calculations for vertical case $\theta=0^\circ$. 
Dashed curves describe the correction factor computed at vertical depth $h$ of 1.15 km for various zenith angles
as a function of slant depth defined by $R=h/\cos\theta$. 
Both solid and dashed curves are shown for four values of cut-off energy 
$E_f$: 10 GeV, 100 GeV, 1 TeV, and 10 TeV, from top to bottom.
\label{fig:cf_bk}}
\end{figure}

The influence of fluctuations of muon losses in matter (mainly due to radiative processes) results in
that the real integral flux $I_{fl}$ is generally greater than $I_{cl}$ calculated in the approximation of continuous losses. 
In this work we propose to allow for the influence of energy loss fluctuations on the value of
angular flux in matter by means of correction factor expressed by the ratio~(\ref{Cf}).   
In the assumption when the same differential cross sections are used for computing both the numerator and denominator
of ratio~(\ref{Cf}), one may expect that this factor depends only weakly on sea level angular spectrum.  
For numerical calculations of the correction factor we used total continuous energy 
losses defined by sum of~(\ref{Ion}),~(\ref{Ionbr}) and~(\ref{Bterm}), and survival probabilities
calculated by using MUM code~\citep{KBS}.  
\begin{table*}
\protect\caption{ Coefficients $c_{ij}$ of the fitting formula~(\protect\ref{CF}) for correction factor calculated for vertical
                  sea level spectrum given by expression~(5) of Ref.~\citep{KBS1}.    
\label{tab:cftab}}
\center{\begin{tabular}{crrrrr} \hline \hline
subscript $i$  & $c_{i0}$~~~~~~~ & $c_{i1}$~~~~~~~ & $c_{i2}$~~~~~~~ & $c_{i3}$~~~~~~~ & $c_{i4}$~~~~~~~ \\\hline
0              &  $ 6.3045 \times10^{-1}$ & $ 6.6658 \times10^{-1}$ & $-4.5138 \times10^{-1}$ & $ 1.2441 \times10^{-1}$ & $-1.1904 \times10^{-2}$ \\ 
1              &  $ 2.0152 \times10^{-1}$ & $-4.2990 \times10^{-1}$ & $ 3.2532 \times10^{-1}$ & $-1.0265 \times10^{-1}$ & $ 1.0751 \times10^{-2}$ \\ 
2              &  $-3.3419 \times10^{-2}$ & $ 5.1833 \times10^{-2}$ & $-3.9229 \times10^{-2}$ & $ 1.2360 \times10^{-2}$ & $-1.2911 \times10^{-3}$ \\ 
3              &  $ 1.6365 \times10^{-3}$ & $-2.3645 \times10^{-3}$ & $ 1.7775 \times10^{-3}$ & $-5.5495 \times10^{-4}$ & $ 5.7557 \times10^{-5}$ \\ 
4              &  $-2.6630 \times10^{-5}$ & $ 3.7770 \times10^{-5}$ & $-2.8207 \times10^{-5}$ & $ 8.7275 \times10^{-6}$ & $-8.9919 \times10^{-7}$ \\ 
\hline \hline
\end{tabular}}
\end{table*} 
The values of correction
factors calculated for the same slant depth $R$ at vertical direction and at zenith angle $\theta$ differ weakly. 
It is illustrated in Fig.~\ref{fig:cf_bk}, where one can see that $C_{f}(\geq E_f,R,0^{\circ})$ differs from  
$C_{f}(\geq E_f,R,\arccos h/R)$ maximum on 3.3$\,\%$ for $E_f>$10 GeV at vertical depth $h$ of 1.15 km. 
It appears that with acceptable accuracy the correction factor depends on slant depth $R$ only, rather than on $R$ and $\theta$
separately. 

The dependencies of correction factor on $E_f$ and $R$, calculated for sea level spectrum
given by expression~(5) of Ref.~\citep{KBS1}, represent the set of rather smooth curves (shown in Fig.~\ref{fig:cf_bk}) 
and it is possible to approximate this factor by formula  
\begin{equation}\label{CF}
C_f(\geq E_f,R,\theta)=\sum_{i=0}^4 ( \sum_{j=0}^4 c_{ij}\log^{j}_{10}\,E_{f} )R^i. 
\end{equation}
Here cut-off energy $E_f$ is expressed in (GeV) and slant depth $R$ is in (km) with the coefficients $c_{ij}$
collected in Table~\ref{tab:cftab}. When using~(\ref{CF}) for cut-off energies $E_f<$10 GeV one should substitute value of $E_f$=10 GeV.

Formula~(\ref{CF}) can be applied for any geometrical shape of the surface. Right hand side of~(\ref{CF}) depends on
$\theta$ because, generally, $R=R(\theta)$. So, in the particular case of a flat surface the angular dependence of the
correction factor appears, in our approximation, only through the relation  
$R=h / \cos\theta$
~(where $h$ is a vertical depth). 

The accuracy of formula~(\ref{CF}) for $E_f$=(1--100)$\,$GeV is better than $\pm2\,\%$ for slant depths $R$ as large as 22 km and
is not worse than $\pm3\,\%$ for $E_f$=1 TeV up to $R$=17 km and for $E_f$=10 TeV up to $R$=15 km.
Fig.~\ref{fig:cf_bk} shows that for $E_f<\,$100 GeV the total energy loss may be treated as quasi-continuous  
(at level of $C_f>\,$0.9) only for slant depths $R<\,$2.5 km but for $E_f$=10 TeV the fluctuations should be taken into account
at level of 15$\,\%$ already for slant depth as small as $R$=1 km. 
For slant depths larger than 10 km $I_{fl}$ and $I_{cl}$ differ more than on a factor of 2.
\balance
\section{Conclusions}
We showed in this paper that the most complicate part of any calculation of muon integral flux at great depths (which 
is taking into account the energy loss fluctuations) can be reduced to a computation of the correction factor
which depends only weakly on the sea level spectrum and is a smooth function of its arguments
(muon cut off energy and a slant depth). The method described here permits to carry out fast analytical
calculations of underwater and underground fluxes, as is shown in Ref.~\citep{KBS1}. These analytical
calculations are greatly facilitated by the parametrizations of all continuous energy losses (using
the most recent expressions for muon interaction cross sections) proposed in the present paper. 

\end{document}